\documentclass[english,aps,tightenlines,showpacs, showkeys, notitlepage]{revtex4-1}
\usepackage[latin9]{inputenc}
\usepackage{amsmath}
\usepackage{amssymb}

\makeatletter
\usepackage{dcolumn}
\usepackage{bm}

\usepackage{babel}

\makeatother

\usepackage{babel}
\begin{document}
\title{Effects of bosonic and fermionic $q$-deformation on the entropic
gravity}
\author{Salih Kibaro\u{g}lu$^{1}$ }
\email{E-mail: salihkibaroglu@gmail.com}

\author{Mustafa Senay$^{1}$}
\email{E-mail: mustafasenay86@hotmail.com}

\date{\today}
\begin{abstract}
In this paper, we study thermodynamical contributions to the theory
of gravity under the $q$-deformed boson and fermion gas models. According
to the Verlinde's proposal, the law of gravity is not based on a fundamental
interaction but it emerges as an entropic force from the changes of
entropy associated with the information on the holographic screen.
In addition, Strominger shows that the extremal quantum black holes
obey neither boson nor fermion statistics, but they obey deformed
statistic. Using these notions, we find $q$-deformed entropy and
temperature functions. We also present the contributions that come
from the $q$-deformed model to the Poisson equation, Newton's law
of gravity and Einstein's field equations. 
\end{abstract}
\affiliation{$^{1}$Department of Natural Sciences, National Defence University,
Istanbul, Turkey }
\maketitle

\section{Introduction}

The mechanism of gravity is described by many methods such as the
General theory of relativity, the Gauge theory of gravity and the
Quantum gravity. These theories try to explain the gravity with different
methods by using geometrical description of spacetime, particles (graviton)
or fields. There is an interesting study published by Jacobson \citep{jacobson1995}
who develop a gravity theory using the first law of thermodynamics
on a local Rindler horizon. This approximation is based on Hawking
and Bekenstein's studies \citep{hawking1971,hawking1975,bekenstein1993}
which are about the relation between the black hole physics and thermodynamics.
In 2011 Verlinde \citep{Verlinde2011} put forward a theory, named
as entropic gravity, which describes gravity as an effect of entropy.
In this theory, Verlinde obtains Newton's second law of gravity and
certain components of the Einstein field equations by using Tolman-Komar
mass and the equipartition rule. The detailed discussion can be found
in \citep{Padmanabhan2010}.

The discovery of the relation between the theory of gravity and thermodynamics
brings a new point of view for understanding the nature of gravity.
Thus we have an alternative theory to get a more general theory of
gravity by deforming thermodynamical quantities such as entropy, temperature
and energy functions. After Verlinde's work this idea become so popular.
Using this notion, the modified Newtonian dynamics (MOND) and generalized
Einstein equation is obtained by deforming energy function with Debye
model \citep{sheykhi2012}. The entropic correction to Coulomb's law
is obtained \citep{hendi2012}. Some developments on modified dark
matter (MDM) is studied in \citep{ho2010,ng2013,edmonds2014}. The
($n+1$) dimensional Einstein field equation is found by \citep{moradpour2015}.
This theory is also applied to quantum black holes \citep{dil2015,dil2016,dil2017,senay2018a}.

On the other hand, according to Strominger's proposal in Ref \citep{strominger1993},
extremal quantum black holes, which have minimum mass and behave like
a particle, obey neither standard Bose nor Fermi statistics. However,
they can be assumed as deformed bosons or fermions and so their thermostatistical
properties can be investigated with the help of deformed Bose or Fermi
statistics. In recent decades, thermodynamical and statistical properties
of deformed boson and fermion systems have been extensively examined
\citep{martin-delgado1991,lavagno2000,gavrilik2012,mirza2011,mohammedzadeh2017,Algin2012,Algin2016,Algin2014,swamy2006}.
Moreover, these deformed boson and fermion systems have a variety
of applications such as understanding of higher-order effects in the
many body interactions \citep{sviratcheva2004}, quantum mechanics
in discontinuous spacetime \citep{dimakis1992}, thermosize effects
\citep{senay2018b}, phonon spectrum in $^{4}He$ \citep{rego-monterio1998},
vortices in superfluid films \citep{bonatsos1992}, and Landau diamagnetism
problem embedded in D-dimensions \citep{brito2011}.

With the above motivations, we consider a $q$-deformed boson and
fermion systems, which previously was studied in \citep{Lavagno2002},
at high temperature regime and obtain generalized entropy and temperature
functions in terms of deformation parameter $q$. Then, we get generalized
forms of the Poisson equation, Newton's second law of gravity and
Einstein's field equations by using $q$-deformed temperature function
based on Verlinde's idea. In the studies \citep{dil2015,senay2018a},
$q$-deformed Einstein equations were obtained by considering different
types $q$-deformed oscillators algebras. The quantum black holes
were assumed as $q$-deformed boson particles in Ref. \citep{dil2015}
and as $q$-deformed fermion particles in Ref \citep{senay2018a}.
However, in this study, we think different $q$-deformed oscillators
algebra than in the Ref. \citep{dil2015,senay2018a} and assume that
the quantum black holes can be described by the $q$-deformed boson
particles or fermion particles. There are several differences between
$q$-deformed model in this study and given in \citep{dil2015,senay2018a}.
For instance, the mean occupation number and the deformed entropy
function in our model have different properties from the Ref. \citep{dil2015,senay2018a}
because of having different quantum oscillators algebra.

The paper is organised as follows. In Section 2, $q$-deformed boson
and fermion gas models are studied at high temperature limit and deformed
entropy function is found. In Section 3, we find generalized Poisson
equation and Newton's second law of gravity by using Verlinde's entropic
gravity proposal and deformed temperature. In Section 4, $q$-deformed
Einstein equation is obtained. The last section concludes the paper
and contains a discussion about possible future application of our
results.

\section{$q$-deformed boson and fermion gas models at h\i gh temperature}

In this section, we briefly introduce the basic algebraic properties
of $q$-deformed boson and fermion gas models. Then, we review some
of the high-temperature thermostatistical properties of such $q$-deformed
boson and fermion models investigated in \citep{Lavagno2002}.

The quantum algebraic structure of the $q$-deformed boson and fermion
is defined by the following relations \citep{Lavagno2002,Ng1990,Lee1990,Chaichian1993,Song1993}

\begin{eqnarray}
 & cc^{*}-\kappa q^{\kappa}c^{*}c=q^{-N},\nonumber \\
 & [\hat{N},c^{*}]=c^{*},\,\,\,\,\,\,[\hat{N},c]=-c,\label{eq: algebra}
\end{eqnarray}
where $c$ and $c^{*}$ are, respectively, deformed annihilation and
creation operator, $\hat{N}$\textit{ }is the total number operator,\textit{
$q$} is the real positive deformation parameter and $\kappa=1$ for
$q$-boson and $\kappa=-1$ for $q$-fermion. Moreover, the operators
obey the relations

\begin{equation}
c^{*}c=\left[\hat{N}\right],\,\,\,\,\,\,c^{*}c=\left[1+\kappa\hat{N}\right],\label{eq: operators}
\end{equation}
where the $q$-basic number is defined as

\begin{equation}
[x]=\frac{q^{x}-q^{-x}}{q-q^{-1}}.\label{eq: q-basic num}
\end{equation}
Also, Jackson derivative (JD) operator for the above mentioned algebra
is given as

\begin{equation}
D^{(q)}f(x)=\frac{1}{x}\left[\frac{f(qx)-f(q^{-1}x)}{q-q^{-1}}\right],
\end{equation}
for any function $f(x)$. This JD operator reduces to the ordinary
derivative operator in the limit $q\rightarrow1$ \citep{Lavagno2002}.

Now, we consider the $q$-deformed boson and fermion gas models constructed
by Eqs. (\ref{eq: algebra})-(\ref{eq: q-basic num}). In the grand
canonical ensemble, the Hamiltonian $H$ of the such a $q$-deformed
gas model have the following form

\begin{equation}
H=\sum_{i}(\varepsilon_{i}-\mu)N_{i},
\end{equation}
where $\varepsilon_{i}$ is the kinetic energy of a particle in the
$i$. state and $\mu$ is the chemical potential. Then, the logarithm
of the grand partition function of the model is given by

\begin{equation}
\ln Z=-\kappa\sum\ln(1-\kappa ze^{-\beta\varepsilon_{i}}),
\end{equation}
where $z=\exp(\mu/kT)$ is the fugacity. Also, this $q$-deformed
gas model has the following mean occupation number

\begin{equation}
N=\sum_{i}n_{i}=\sum_{i}\frac{1}{q-q^{-1}}\ln\left(\frac{z^{-1}e^{\beta\varepsilon_{i}}-\kappa q^{-\kappa}}{z^{-1}e^{\beta\varepsilon_{_{i}}}-\kappa q^{\kappa}}\right).
\end{equation}
In order to obtain the high-temperature thermostatistical properties
of the model, the sum over states can be replaced with the integral
for a large volume and a large number of particles. Therefore, the
equation of state $PV/kT=\ln Z$ and the particle density can be written,
respectively, as

\begin{equation}
\frac{P}{kT}=\frac{1}{\lambda^{3}}\frac{4}{3\sqrt{\pi}}\intop_{0}^{\infty}x^{3/2}dx\,\frac{1}{q-q^{-1}}\ln\left(\frac{1-\kappa q^{-\kappa}ze^{-x}}{1-\kappa q^{\kappa}ze^{-x}}\right),
\end{equation}

\begin{equation}
\frac{N}{V}=\frac{1}{\lambda^{3}}\frac{2}{\sqrt{\pi}}\intop_{0}^{\infty}x^{1/2}dx\,\frac{1}{q-q^{-1}}\ln\left(\frac{1-\kappa q^{-\kappa}ze^{-x}}{1-\kappa q^{\kappa}ze^{-x}}\right),
\end{equation}
where $\lambda=h/\sqrt{2\pi mkT}$ is the thermal wavelength, $x=\beta\varepsilon$
and $\varepsilon=p^{2}/2m$. For high-temperature limits $z\ll1$,
these integrals can be expanded with Taylor series and defined as

\begin{equation}
\frac{P}{kT}=\frac{1}{\lambda^{3}}h_{5/2}^{\kappa}(z,q),\label{eq: taylor1}
\end{equation}

\begin{equation}
\frac{N}{V}=\frac{1}{\lambda^{3}}h_{3/2}^{\kappa}(z,q),\label{eq: taylor2}
\end{equation}
where \textit{q}-deformed $h_{n}^{\kappa}(z,q)$ is defined as

\noindent 
\begin{align}
h_{n}^{\kappa}(z,q) & =\frac{1}{\Gamma(n)}\intop_{0}^{\infty}x^{n-1}dx\,\frac{1}{q-q^{-1}}\ln\left(\frac{1-\kappa q^{-\kappa}ze^{-x}}{1-\kappa q^{\kappa}ze^{-x}}\right)\nonumber \\
 & =\frac{1}{q-q^{-1}}\left[\sum_{l=1}^{\infty}\frac{(\kappa zq^{\kappa})^{l}}{l^{n+1}}-\sum_{l=1}^{\infty}\frac{(\kappa zq^{-\kappa})^{l}}{l^{n+1}}\right].
\end{align}
In the limit $q\rightarrow1$, the deformed $h_{n}^{\kappa}(z,q)$
functions reduce to the standard Bose-Einstein functions $g_{n}(z)$
for bosons and Fermi-Dirac functions $f_{n}(z)$ for fermions \citep{Patriha}.
The internal energy of the model can be obtained by using the thermodynamical
relation $U=-(\partial\ln Z/\partial\beta)$$_{z,V}$ as

\begin{equation}
U=\frac{3}{2}\frac{kTV}{\lambda^{3}}h_{5/2}^{\kappa}(z,q).\label{eq: internal en}
\end{equation}
Now, we want to determine the Helmholtz free energy of the model from
the thermodynamical relation $A=\mu N-PV$. From Eqs. (\ref{eq: taylor1})
and (\ref{eq: taylor2}), the Helmholtz free energy of the model can
be derived as

\begin{equation}
A=\frac{kTV}{\lambda^{3}}\left[h_{3/2}^{\kappa}(z,q)\ln z-h_{5/2}^{\kappa}(z,q)\right].\label{eq: helmholtz en}
\end{equation}
Then, the $q$-deformed entropy function of the model can be found
from the thermodynamical relation $S=(U-A)/T$. From Eqs. (\ref{eq: internal en})
and (\ref{eq: helmholtz en}), the entropy of the model becomes

\begin{equation}
S=\frac{kV}{\lambda^{3}}\left[\frac{5}{2}h_{5/2}^{\kappa}(z,q)-h_{3/2}^{\kappa}(z,q)\,\ln z\right].
\end{equation}
This deformed entropy function of the model can also be written in
the following form

\begin{equation}
S=\frac{(2\pi m)^{3/2}V}{Th^{3}}E^{5/2}H^{\kappa}(z,q),\label{eq: def. S}
\end{equation}
where $E=kT$ is the one-particle average kinetic energy and $H^{\kappa}(z,q)$
is defined as

\begin{equation}
H^{\kappa}(z,q)=\frac{5}{2}h_{5/2}^{\kappa}(z,q)-h_{3/2}^{\kappa}(z,q)\ln z.
\end{equation}
In the next section, we will particularly focus on the derivation
of the Poisson equation for gravity from the $q$-deformed entropy
function defined in\emph{ }\textit{\emph{Eq.(\ref{eq: def. S}) by
taking account of the Verlinde's entropic gravity approach.}}

\section{modified po\i sson equat\i on and newton's law of gravitation}

According to the Verlinde's approach \citep{Verlinde2011}, the theory
of gravity (including Newton's law of gravity and General theory of
relativity) can be obtained by using the holographic principle when
the mass is distributed over a holographic screen. In this section,
we investigate the effects of the $q$-deformed model on the Poisson
equation and Newton's law of gravity. For this purpose, we firstly
derive deformed temperature by using deformed entropy function given
in Eq.(\ref{eq: def. S}) with the methods presented in \citep{senay2018a}.
The total entropy $S$ remains constant when the entropic force is
equal to the force increasing the entropy. In this case, the system
reaches the statistical equilibrium and the variation of the entropy
goes to zero, such that

\begin{equation}
\frac{d}{dx^{a}}S(E,x^{a})=0,\label{eq: var ent}
\end{equation}
and it can be written as following relation

\begin{equation}
\frac{\partial S}{\partial E}\frac{\partial E}{\partial x^{a}}+\frac{\partial S}{\partial x^{a}}=0.\label{eq: var ent 2}
\end{equation}

To obtain deformed temperature, we now assume that the mass enclosed
by the surface is formed by the $q$-deformed bosons or fermions.
Accordingly, the $q$-deformed entropy function in Eq.(\ref{eq: def. S})
can be used to find the deformed temperature. So that the Eq.(\ref{eq: var ent 2})
gives

\begin{equation}
-\frac{5V(2\pi mE)^{3/2}}{2h^{3}}H^{\kappa}(z,q)F_{a}+T\nabla_{a}S=0,\label{eq: def_temp_1}
\end{equation}
where $\partial E/\partial x_{a}=-F_{a}$ and $\partial S/\partial x_{a}=\nabla_{a}S$.
By applying the relations $F=ma=-m\nabla\Phi$ and $\nabla_{a}S=(-2\pi mN_{a})/\hbar$
on the Eq.(\ref{eq: def_temp_1}). Also, the deformed temperature
on the holographic screen $\mathcal{S}$ can be obtained as

\begin{eqnarray}
T & = & \frac{5V\left(2mE\right)^{3/2}}{8\sqrt{\pi}h^{2}}H^{\kappa}\left(z,q\right)N^{a}\nabla_{a}\Phi,\nonumber \\
 & = & \tilde{\alpha}\left(z,q\right)T_{U},\label{eq:def_temp}
\end{eqnarray}
where $T_{U}=\frac{\hbar}{2\pi}N^{a}\nabla_{a}\Phi$ represents the
standard Unruh temperature and $N_{a}$ is the unit outward pointing
vector which is normal to the screen. The parameter $\tilde{\alpha}\left(z,q\right)$
is defined as follows,

\begin{equation}
\tilde{\alpha}\left(z,q\right):=\frac{5V\left(2\pi mE\right)^{3/2}}{2h^{3}}H^{\kappa}\left(z,q\right).\label{eq: alpha}
\end{equation}

Let us now derive a modified version of the Poisson equation based
on the entropic gravity approach by using the $q$-deformed temperature
obtained in Eq. (\ref{eq:def_temp}). According to Bekenstein \citep{bekenstein1993},
if we suppose that there is a test particle near the black hole horizon
which is distant from one Compton wavelength, it increases the black
hole mass and horizon area. This process is identified as one bit
of information. According to the holographic principle, the total
number of bits $N$ is proportional to the area $A$,

\begin{equation}
N=\frac{A}{G\hbar},\label{eq: num_of_bits}
\end{equation}
where the speed of light $c=1$ and $G$ is a constant. The constant
$G$ will be identified with Newton's gravitational constant when
we consider gravity. Now, we suppose that the total energy of the
system $E$ is associated with the total mass distributed over all
the bits. By taking into account the equipartition law of energy,
the total mass can be defined \citep{Verlinde2011} as

\begin{equation}
M=\frac{1}{2}\int_{\mathcal{S}}TdN.\label{eq: mass_total}
\end{equation}
where the integration over the holographic screen $\mathcal{S}$.
Substituting Eq. (\ref{eq:def_temp}) in Eq.(\ref{eq: mass_total}),
then using Eq.(\ref{eq: num_of_bits}) for the number of bits on the
holographic screen Eq.(\ref{eq: num_of_bits}), we obtain

\begin{equation}
M=\frac{\tilde{\alpha}\left(z,q\right)}{4\pi G}\intop_{\mathcal{S}}\nabla\Phi dA.\label{eq: mass_2}
\end{equation}
Using the divergence theorem, the Eq.(\ref{eq: mass_2}) can be written
as

\begin{equation}
M=\frac{\tilde{\alpha}\left(z,q\right)}{4\pi G}\intop_{V}\nabla\cdot(\nabla\Phi)dV,\label{eq: mass3}
\end{equation}
where $V$ represents three dimensional volume element. On the other
hand, the mass distribution in the closed surface can be given as

\begin{equation}
M=\intop_{V}\rho(r)dV,\label{eq: mass4}
\end{equation}
where $\rho(r)$ is the mass density. Comparing Eq.(\ref{eq: mass3})
with Eq.(\ref{eq: mass4}), we get the Poisson equation for gravity
for $q$-deformed boson and fermion models as follows,

\begin{equation}
\nabla^{2}\Phi(r)=4\pi G_{eff}\rho(r),\label{eq: poisson1}
\end{equation}
where $G_{eff}$ is effective or $q$-deformed Newton's gravitational
constant and defined as,

\begin{equation}
G_{eff}:=G\tilde{\alpha}^{-1}\left(z,q\right).\label{eq: deformed_G}
\end{equation}

Now if we define the Newton potential and gravitational acceleration
as follows respectively, 
\begin{equation}
\Phi\left(r\right):=-\frac{G_{eff}M}{r},\label{eq: potential_newton}
\end{equation}

\begin{equation}
\mathbf{g}=-\mathbf{\nabla}\Phi\left(r\right),\label{eq: gravitational_acce}
\end{equation}
then the Eq. (\ref{eq: poisson1}) takes following form, 
\begin{equation}
\nabla\cdot\mathbf{g}=-4\pi G_{eff}\rho(r).\label{eq: poisson2}
\end{equation}

After these calculations, substituting Eqs. (\ref{eq: potential_newton})
and (\ref{eq: gravitational_acce}) in Newton's second law $\mathbf{F}=m\mathbf{g}$
we find $q$-deformed force formula with effective gravitational constant
as,

\begin{equation}
F=-\frac{G_{eff}Mm}{r^{2}}.\label{eq: def_newton}
\end{equation}

The Eqs. (\ref{eq: poisson1}) and (\ref{eq: def_newton}) show that
the gravity can be modified by $q$-deformed temperature function
based on Verlinde's proposal. In other words, the Eq.(\ref{eq: def_newton})
can be seen as the generalization of the Newton's law of universal
gravitation.

\section{generalization of the einstein equations}

Here, we focus on generalizing the Einstein equatiosn by applying
Verlinde's entropic gravity proposal \citep{Verlinde2011}. For this
aim, we will use a deformed temperature. Before calculating the Einstein
equations we give some information about the deformed temperature
function in Eq.(\ref{eq:def_temp}). The Unruh temperature on the
holographic screen is given as

\begin{equation}
T_{U}=\frac{\hbar a}{2\pi},
\end{equation}
where the constants $c$ and $k_{B}$ equal to one and $a$ represents
the acceleration which is perpendicular to screen $\mathcal{S}$.
This temperature is experienced by an observer in an accelerated frame.
The acceleration also defined with the Newton potential $\phi$ in
general relativity as

\begin{equation}
a^{b}=-\nabla^{b}\phi.
\end{equation}
Here $\phi$ can be seen as a generalization of Newton potential in
general relativity when we describe it by time like killing vector
$\xi^{a}$ \citep{Wald1984},

\begin{equation}
\phi=\ln\left(-\xi^{a}\xi_{a}\right).
\end{equation}
The exponential $e^{\phi}$ represents the redshift factor that relates
local time coordinate to that at a reference point with $\phi=0$,
which we will take to be at infinity \citep{Verlinde2011}. When taking
into account the redshift factor $e^{\phi}$, the Unruh temperature
$T_{U}$ on the holographic screen as a equipotential surface with
non relativistic case is defined as,

\begin{equation}
T_{U}=\frac{\hbar}{2\pi}e^{\phi}N^{a}\nabla_{a}\phi.
\end{equation}

In a similar calculations given in previous section with generalized
Newton's potential, the deformed temperature can be found as,

\begin{eqnarray}
T & = & \frac{5V}{8\sqrt{\pi}}\frac{\left(2mE\right)^{3/2}}{h^{2}}H^{\kappa}\left(z,q\right)e^{\phi}N^{a}\nabla_{a}\phi,\nonumber \\
 & = & \tilde{\alpha}\left(z,q\right)T_{U}.\label{eq: def_temp_red}
\end{eqnarray}
Substituting Eq.(\ref{eq: def_temp_red}) in Eq.(\ref{eq: mass_total})
and using Eq.(\ref{eq: num_of_bits}), we obtain for the total mass

\begin{equation}
M=\frac{\tilde{\alpha}\left(z,q\right)}{4\pi G}\intop_{\mathcal{S}}e^{\phi}\nabla\phi dA.\label{eq: mass_total_2}
\end{equation}
where $\tilde{\alpha}\left(z,q\right)$ is defined in Eq.(\ref{eq: alpha}).
The Eq.(\ref{eq: mass_total_2}) can be seen as the natural generalization
of Gauss's law to the General Relativity for $q$-deformed model and
the mass $M$ corresponds to Komar mass. The Komar mass can also be,
alternatively, re-expressed in terms of the Killing vector $\xi^{a}$
and Eq. (\ref{eq: deformed_G}) as,

\begin{equation}
M=\frac{1}{4\pi G_{eff}}\intop_{\mathcal{S}}dx^{a}\wedge dx^{b}\epsilon_{abcd}\nabla^{c}\xi^{d}.\label{eq: mass5}
\end{equation}

According to the Stokes theorem, the surface integral in Eq.(\ref{eq: mass5})
can be converted into a volume integral if the surface is two dimensional
boundary of the hyper-surface. Using the relation $\nabla_{a}\nabla^{a}\xi^{b}=-R_{\,\,\,a}^{b}\xi^{a}$,
which is implied by the Killing equation for $\xi^{a}$, the Komar
mass can be written as

\begin{equation}
M=\frac{1}{4\pi G_{eff}}\intop_{\varSigma}R_{ab}n^{a}\xi^{b}dV,\label{eq: mass6}
\end{equation}
where $R_{ab}$ is the Ricci curvature tensor, $\Sigma$ is the three
dimensional volume bounded by the holographic screen and $n^{a}$
is its outward normal. Furthermore, the Komar mass can also be written
as an integral over the enclosed volume of certain components of the
energy-momentum tensor $T_{ab}$ \citep{Wald1984} as 
\begin{eqnarray}
M & = & 2\int_{\Sigma}\left(T_{ab}-\frac{1}{2}g_{ab}T\right)n^{a}\xi^{b}dV,\label{eq: mass7}
\end{eqnarray}
where $g_{ab}$ is space time metric tensor. When Eqs. (\ref{eq: mass6})
and (\ref{eq: mass7}) are equated, we find

\begin{equation}
R_{ab}=8\pi G_{eff}\left(T_{ab}-\frac{1}{2}g_{ab}T\right)\label{eq: einstein1}
\end{equation}
This gives us only a time-time component of the Einstein field equations
\citep{Verlinde2011,li2012}. Taking the trace of Eq.(\ref{eq: einstein1})
leads to

\begin{equation}
R_{ab}-\frac{1}{2}g_{ab}R=8\pi G_{eff}T_{ab}.\label{eq: einstein2}
\end{equation}

The last equation is the $q$-deformed Einstein equations resulting
from considering the quantum black holes as deformed bosons or fermions.
The factor $\tilde{\alpha}\left(z,q\right)$ in the definition of
$G_{eff}$ Eq.(\ref{eq: deformed_G}) carries the information of the
total energy-momentum of the deformed systems.

\section{Conclusion}

In this study, we have derived deformed entropy and temperature functions
for $q$-deformed boson and fermion gas models at high temperature
limit. Using Verlinde's entropic gravity approach with deformed temperature
function, we have obtained $q$-deformed Poisson's equations for gravity,
Newton's law of universal gravitation and the Einstein field equations.
According to the Strominger's idea, quantum black holes obey deformed
statistics, neither boson nor fermion. Thus our results provide an
alternative framework for describing gravitational effect around the
quantum black holes.

In the limit $q\rightarrow1$ the deformed boson and fermion gas models
transform into its ideal form in which there is no interaction between
related particles, in other words, the model goes non-deformed case.
Thus, we can not use this model to describe the extremal black holes
in this limit because it needs deformed statistic. Furthermore, the
deformed temperature functions of the model, given in Eq.(\ref{eq:def_temp})
and Eq.(\ref{eq: def_temp_red}), do not reduce to their standard
forms in $q\rightarrow1$ limit because the parameter $\tilde{\alpha}\left(z,q\right)$
can not be equal to one. Due to the effective gravitational constant
Eq.(\ref{eq: deformed_G}) which depend on $\tilde{\alpha}\left(z,q\right)$,
our deformed gravitational equations Eq.(\ref{eq: poisson2}), Eq.(\ref{eq: def_newton})
and Eq.(\ref{eq: einstein2}) remain deformed under this limit.

Here we want to give a remark about possible application. If we consider
the Brans-Dicke theory of gravity \citep{brans1961,dirac1973,charap1974},
the Lagrangian density which is invariant under the scale transformation
can be written as

\begin{equation}
\mathcal{L}=\frac{1}{16\pi}\intop d^{4}x\sqrt{-g}\varphi^{2}R\label{eq: Lagrangian}
\end{equation}
where $g$ is determinant of metric tensor, $R$ is the Ricci scalar
and $\varphi\left(x\right)$ is a scalar field. In this case, the
Einstein equations take the following form:

\begin{equation}
R_{ab}-\frac{1}{2}g_{ab}R=8\pi\varphi^{-1}T_{ab},\label{eq: einstein eq - BD}
\end{equation}
and comparing the Eqs. (\ref{eq: einstein2}) and (\ref{eq: einstein eq - BD})
a close connection emerges between $\varphi\left(x\right)$ and $\tilde{\alpha}\left(z,q\right)$
as,

\begin{equation}
\varphi\left(x\right)\thicksim\frac{\tilde{\alpha}\left(z,q\right)}{G}=\frac{1}{G_{eff}}
\end{equation}
this relation means that the scalar field $\varphi\left(x\right)$
can be described by thermodynamical quantities in the deformed model
and thus, if it is correct, one can say that the Brans-Dicke like
theory of gravity can be obtained by using Verlinde's entropic gravity
approach under the $q$-deformed systems.

\end{document}